\documentclass[a4paper,12pt]{article}
\usepackage{epsfig}
\usepackage{graphicx}
\usepackage{dcolumn}
\usepackage{bm}
\usepackage{longtable}
\usepackage{amssymb}
\usepackage{amsfonts}
\usepackage[margin=1.0in]{geometry}
\usepackage{hyperref}
\usepackage{amsmath}
\usepackage{multirow}
\allowdisplaybreaks[1]

\def\dd{\textrm{d}}  				     

\newcommand{\be}{\begin{equation}}
\newcommand{\ee}{\end{equation}}
\newcommand{\one}{{\rm 1\kern -.9mm l}} 

\usepackage{slashed}

\def\bB{\mathbf{B}}

\def\cD{ {\cal D} }

\makeatletter							  
\@addtoreset{equation}{section}					  
\makeatother							  

\begin{document}
\begin{titlepage}
\begin{flushright}
DFT 10/2012\\
DISIT-2012\\
DFPD-12/TH/10

\par\end{flushright}
\vskip 1.5cm
\begin{center}
\textbf{\huge \bf Fermionic Wigs for BTZ Black Holes}
\textbf{\vspace{2cm}}\\
{\Large L.G.C.~Gentile$ ^{~a, c, e,}$\footnote{lgentile@pd.infn.it},$\ $ P.A.~Grassi$ ^{~a, d,}$\footnote{pgrassi@mfn.unipmn.it}  and A.~Mezzalira$ ^{~b, d,}$\footnote{mezzalir@to.infn.it}}

\begin{center}
{a) { \it DISIT, Universit\`{a} del Piemonte Orientale,
}}\\
{{ \it via T. Michel, 11, Alessandria, 15120, Italy, }}
 \\ \vspace{.2cm}
 {b) { \it Dipartimento di Fisica Teorica, Universit\`a di Torino,}}\\
 {{\it via P. Giuria, 1, Torino, 10125, Italy,}}
\\ \vspace{.2cm}
{c) { \it Dipartimento di Fisica Galileo Galilei,\\
Universit\`a di Padova,\\
via Marzolo 8, 35131 Padova, Italy,
}}
 \\ \vspace{.2cm}
 {d) { \it INFN - Gruppo Collegato di Alessandria - Sezione di Torino,}}
 \\ \vspace{.2cm}
 {e) { \it INFN, Sezione di Padova,\\
via Marzolo 8, 35131, Padova, Italy.}}
\end{center}

\par\end{center}
\vfill{}

\begin{abstract}
{\vspace{.3cm}

       \noindent
We compute the wig for the BTZ black hole, namely
the complete non--linear solution of supergravity equations with all fermionic zero modes. 
We use a ``gauge completion'' method starting from $AdS_{3}$ Killing spinors
to generate the gravitinos fields associated to the BH and we compute the back--reaction on the metric.
Due to the anticommutative properties of the fermionic hairs the resummation of these effects
truncates at some order. 
We illustrate the technique proposed in a precedent 
paper in a very explicit and analytical form. 
We also compute the mass, the angular momentum and other charges with their corrections.

}
\end{abstract}
\vfill{}
\vspace{1.5cm}
\end{titlepage}

\vfill
\eject

\setcounter{footnote}{0}


\section{Introduction}

In \cite{Gentile:2012jm} we constructed the complete solution (named ``{\it wig}" to recall that 
all fermionic hairs are resummed into a complete non--linear solution)  starting from non-extremal black holes for $\mathcal{N}=2$, $D=5$ and $D=4$ 
supergravity. The construction presented in that work was based on a \verb Mathematica \textsuperscript{\textregistered} package used to ``integrate'' the 
fermionic zero modes of the solution into the metric and the other fields. 
The word ``integrate'' means, in the present context, the resummation of all fermionic contributions to fundamental fields.
This procedure is also known as ``gauge completion'' and it has been used to resum all components of a given superfield.

In the present paper we analyze a simpler situation where it possible to compute all contributions analytically and presenting them in a compact and manageable form.
For that we consider $\mathcal{N}=2$, $D=3$ supergravity theory \cite{Achucarro:1987vz,Banados:1992wn,Banados:1992gq,Coussaert:1993jp,Izquierdo:1994jz,Howe:1995zm,Banados:1998pi,Henneaux:1999ib}.

It has been pointed out that this theory is a topological one and therefore it does not possess any local degrees of freedom,
that is all fluctuations can be reabsorbed by gauge redefinitions.
Nevertheless the theory has non--trivial localized solutions with singularities such as black holes, which are trivial solution except for the fixed points of an orbifold action (the orbifold is defined in terms of a discrete subgroup of the isometry group \cite{Banados:1992gq}).

Our motivations stem from the fluid/gravity correspondence discovered in
\cite{Bhattacharyya:2008jc,Rangamani:2009xk} and extended in  \cite{Gentile:2012jm} to fermionic degrees of freedom. Starting from a 
solution of a gravity/supergravity theory on $AdS$ background  such as black hole or 
black-brane solution, one acts with certain isometries transformations whose parameters depend on $AdS$--boundary coordinates.
The transformed expressions
are no longer solution of the equations of motion unless those local parameters 
satisfy some non-linear differential  equations. 
They are the Navier-Stokes equations for the boundary field theory. In 
\cite{Gentile:2011jt}, we showed that by extending the construction of \cite{Bhattacharyya:2008jc,Rangamani:2009xk} it is possibile to 
derive the fermionic corrections to Navier-Stokes equations in terms of fermion bilinears. The latter may acquire a non-vanishing 
expectation value yielding physical modifications of the fluid dynamics. Despite the originality of the result, our analisys in \cite{Gentile:2011jt} 
was limited to the linear approximation since we did not possess the result after a finite superisometry. 
For that reason,
in \cite{Gentile:2012jm} we construct the general solution. That 
 reveals several interesting aspects that we present here in a simpler set--up. 

The analysis in \cite{Gentile:2012jm} is based on the papers by Aichelburg and Embacher \cite{Aichelburg:1986wv,Aichelburg:1987hy,Aichelburg:1987hx,Aichelburg:1987hz,Aichelburg:1987ia} where the Schwarzschild solution for $\mathcal{N}=2$, $D=4$ supergravity in flat space 
has been lifted to a full non--linear solution of the equations of motion including the fermionic zero modes. 
The fields are constructed iteratively starting from the pure bosonic expressions and acting on them with the supersymmetry. 
In \cite{Gentile:2012jm}, we adopt their construction starting from non--extremal black holes in $\mathcal{N}=2$, $D=5$ supergravity in $AdS_5$ background.
The procedure can be also adapted to BPS 
objects and it will be presented in forthcoming publications \cite{INPREP}. 
The choice of the non-extremal solution is made to 
simplify the computation since all supersymmetries are broken and any transformation can be used to construct the complete wig. 

In the present work we start from BTZ black holes \cite{Banados:1992wn,Banados:1992gq,Coussaert:1993jp} 
for $\mathcal{N}=2$, $D=3$ supergravity \cite{Izquierdo:1994jz,Howe:1995zm}
and we construct the corresponding wig. 
We first compute the fermionic zero modes then in terms of them the gravitinos, partners of the black hole, and finally the complete solution.
The gauge field, which is 
zero at the bosonic level, becomes non--zero by fermionic corrections.
We compute all charges associated to 
the BTZ black hole with all fermionic contributions.
Finally, we compute the new stress--energy tensor on the boundary of $AdS$ 
relevant for the fluid/gravity correspondence.

\section{$\mathcal{N}=2$, $D=3$ $AdS$ Supergravity}

As mentioned in the introduction, we consider the simplest non--trivial example of $\mathcal{N}=2$, $D=3$ of \cite{Achucarro:1987vz,Izquierdo:1994jz}
which is described by the vielbein $e^{A}$, the gravitino (complex) $\psi$, an abelian gauge field $A$ and the spin connection $\omega^{AB}$.
Those are the gauge fields of the diffeomorphism, the local supersymmetry, the local $U\left( 1 \right)$ transformations and of Lorentz symmetry.
The gauge symmetry can be used to gauge away all local degrees of freedom except when some fixed points are present, 
namely for localised singular solutions \cite{Banados:1992wn,Banados:1992gq,Henneaux:1999ib,Henneaux:1984ei}

The invariant action has the following form
\begin{align}
	S = & \int \left( 
		R^{AB} \wedge e^{C} \varepsilon_{ABC}	
		-
		\frac{\Lambda}{3} e^{A} \wedge e^{B} \wedge e^{C} \varepsilon_{ABC}
		-
		\bar \psi \wedge {\cal D} \psi
		-
		A \wedge \dd A
		\right)
		\ ,
	\label{action0}
\end{align} 
which in components reads
\begin{align}
	S= \int\! d^3x\, \left[
		e \left( R + 2 \Lambda \right) 
		- \bar\psi_{M} {\cal D}_{N} \psi_{R} \varepsilon^{MNR}
		- A_{M} \partial_{N} A_{R} \varepsilon^{MNR}
	\right]\ ,
\label{action1}
\end{align}
where $R$ is the Ricci scalar, $\left\{ A,B,\dots \right\}$ label flat indices and $\left\{ M,N,\dots \right\}$ refer to curved ones.
The action is invariant under all gauge transformations and it can be cast in a Chern--Simons form. 
For $AdS_{3}$, the cosmological constant is $\Lambda=-1$.\footnote{Note that $AdS_{3}$ radius is set to one.} 
The covariant derivative ${\cal D}$ is defined as
\begin{align}
	{\cal D}_M = D_M + \frac{i}{2} A_M + {1\over2}\Gamma_M \ ,
\label{action2}
\end{align}
where $D= \dd + \frac{1}{4}\omega^{AB} \Gamma_{AB}$ is the usual Lorentz-covariant differential.
It can be easily shown that (\ref{action1}) is invariant under the supersymmetry transformations
\begin{align}
\delta_\epsilon\psi &= {\cal D}\epsilon \ ,&
\delta_\epsilon e^A &= {1\over4}\left( \bar\epsilon\Gamma^A\psi - \bar\psi\Gamma^A\epsilon \right)
\ ,&
\delta_\epsilon A &= {i\over 4} \left( \bar\epsilon \psi - \bar\psi\epsilon \right) \ .
\label{action4}
\end{align}
The spin connection transforms accordingly when the vielbein postulate is used to compute $\omega^{AB}$.
The metric signature is
$(-,+,+)$ and the gamma
matrices $\Gamma^A$ are real. From (\ref{action1}) we deduce the following equations of motion
\begin{align}
	& \cD \psi = 0 
	\ , \nonumber \\
	& \dd A = \frac{i}{4} \bar \psi \wedge \psi
	\ , \nonumber \\
	& D e^{A} = {1\over4}\bar\psi\wedge\Gamma^A \psi 
	\ ,\nonumber \\	
	& R^{AB} - \Lambda e^{A} \wedge e^{B} = \frac{1}{4} \varepsilon^{AB}{}_{C} \bar\psi\Gamma^{C}\psi 
	\ .
	\label{EoMs1}
\end{align}
The third equation is the vielbein postulate, from which the spin connection $\omega^{AB}$ is computed.
It is possible to check the above equations against the Bianchi identities. 
Note that the theory is topological and therefore it can be written in the form language.\footnote{
Using the forms, the gauge symmetries are obtained by shifting all fields 
 $e^{A}\rightarrow e^A + \xi^{A}$, $\psi\rightarrow\psi+\eta$, $\omega\rightarrow\omega^{AB} + k^{AB}$ and $A\rightarrow A+C$
and consequently the differential operator $\dd\rightarrow \dd+s$. $\xi^{A}$, $\eta$, $k^{AB}$ and $C$ are the ghosts associated to diffeomorphism, supersymmetry, Lorentz symmetry and $U\left( 1 \right)$ transformation, respectively and $s$ is the BRST differential associated to those gauge symmetries.}
The gravitino equation is nothing else than the vanishing of its field strength, 
the second one fixes the field strength of the gauge field and the fourth one fixes the Riemann tensor.

\section{$AdS_{3}$ and BTZ Black Hole}

In the present section we describe two solutions of supergravity equations of motion (\ref{EoMs1}): the $AdS_{3}$ space and the BTZ black hole in $AdS_{3}$.

In global coordinates, $AdS_{3}$ metric is 
\begin{align}
	\dd s^{2} =
	- f^{2} \dd t^{2} 
	+ f^{-2} \dd r^{2}
	+ r^{2} \dd \phi^{2}
	\ ,
	\label{AdSCglobal1}
\end{align}
where $f^{2}= r^{2} + 1$. 
The associated vielbeins are
\begin{align}
	  e^{0} = &\  f \dd t \ ,
	& e^{1} = &\  f^{-1} \dd r \ ,
	& e^{2} = &\  r \dd \phi 
	\ ,
\label{vielb1}
\end{align}
and the spin connection components read
\begin{align}
	\omega^0{}_{1} = &\ r \dd t \ ,
	& \omega^{0}{}_{2} = &\  0 \ ,
	& \omega^2{}_{1} = &\  f \dd \phi \ .
	\label{spinC1}
\end{align}
The flat metric $\eta_{AB}$ is mostly plus  $\left( -,+,+ \right)$.
The gamma matrices are defined as follows
\begin{align}
	\Gamma_{0} = &\  i \sigma_{2}
	\ , &
	\Gamma_{1} = &\  \sigma_{3}
	\ , &
	\Gamma_{2} = &\  \sigma_{1}
	\ , &
	\left\{ \Gamma_{A}\,,\,\Gamma_{B} \right\} = & 2 \eta_{AB}
	\ .
	\label{gammaBTZ}
\end{align}
Then, the Killing spinor equations read
\begin{align}
	& \partial_{r} \, \epsilon + \frac{1}{2 f} \Gamma_{1} \, \epsilon	= 0 \ ,
	\nonumber\\
	& 
	\partial_{t} \, \epsilon
	+
	\frac{1}{2} \left( - r \Gamma_{2}  + f \Gamma_{0} \right) \, \epsilon = 0 \ ,
	\nonumber\\
	& 
	\partial_{\phi} \, \epsilon
	+
	\frac{1}{2}  \left( r \Gamma_{2}  - f \Gamma_{0} \right) \, \epsilon = 0 \ .
	\label{KilEq2BTZ}
\end{align}
The index of gamma matrices is flat since the vielbein is written explicitly.
Note that $\left( f \Gamma_{1} + r \Gamma_{0} \right)^{2} = \one$. 
To solve eqs.~(\ref{KilEq2BTZ}), we define the projected spinors
\begin{align}
	\epsilon_{\pm} = \pm \Gamma_{1} \epsilon_{\pm}
	\ ,
	\label{KilEq3}
\end{align}
hence, equations (\ref{KilEq2BTZ}) read
\begin{align}
	&
	\partial_{r} \epsilon_{+} + \frac{1}{2 f} \epsilon_{+} =  0
	\ ,
	&
	\partial_{r} \epsilon_{-} - \frac{1}{2 f} \epsilon_{-} =  0
	\ , & \nonumber \\
	&
	\partial_{t} \epsilon_{+} + \frac{1}{2} \left( f - r \right) \epsilon_{-} =  0
	\ ,
	&
	\partial_{t} \epsilon_{-} - \frac{1}{2} \left( f + r \right) \epsilon_{+} =  0
	\ , & \nonumber \\
	&
	\partial_{\phi} \epsilon_{+} + \frac{1}{2} \left( r - f \right) \epsilon_{-} =  0
	\ ,
	&
	\partial_{\phi} \epsilon_{-} + \frac{1}{2} \left( r + f \right) \epsilon_{+} =  0
	\ .
	\label{eps2BTZ}
\end{align}
Solving the $r$--equations we have
\begin{align}
	\epsilon_{+} = & \left( r + f \right)^{-1/2} \eta_{+}\left( t,\phi \right) 
	\ , 
	& \epsilon_{-} = & \left( r + f \right)^{1/2}  \eta_{-}\left( t,\phi \right)
	\ ,
	\label{eps21BTZ}
\end{align}
thus the $t$--and $\phi$--equations reduce to
\begin{align}
	\partial_t \eta_+ + \frac{1}{2} \eta_- = & 0
	\ ,
	&
	\partial_t \eta_- - \frac{1}{2} \eta_+ = & 0
	\ ,\nonumber\\
	\partial_\phi \eta_+ - \frac{1}{2} \eta_- = & 0
	\ ,
	&
	\partial_\phi \eta_- + \frac{1}{2} \eta_+ = & 0
	\ .
	\label{eps22BTZ}
\end{align}
The solutions read
\begin{align}
	\epsilon_{+} = & \left( r + f \right)^{-1/2} \left( \zeta_{1} \cos \left[ \frac{t - \phi}{2} \right] - \zeta_{2} \sin \left[ \frac{t - \phi}{2} \right]  \right)
	\ , \nonumber \\
	\epsilon_{-} = & \left( r + f \right)^{1/2} \left( \zeta_{2} \cos \left[ \frac{t - \phi}{2} \right] + \zeta_{1} \sin \left[ \frac{t - \phi}{2} \right]  \right)
	\ ,
	\label{eps3BTZ}
\end{align}
that is
\begin{align}
	\epsilon
	= &
	\frac{1}{2} 
	\left[
	\left( \sqrt{r + f} + \frac{1}{\sqrt{r + f}}  \right) \one
	-
	\left( \sqrt{r + f} - \frac{1}{\sqrt{r + f}}  \right) \Gamma_{1}	
	\right]\times
	\nonumber\\
	&\ \times
	\left( 
	\cos \left[ \frac{t - \phi}{2} \right] \one 
	-
	\sin \left[ \frac{t - \phi}{2} \right] \Gamma_{0}
	\right) \zeta
	\ ,
	\label{eps4BTZ}
\end{align}
where $\zeta$ is a spinor with two complex Grassmann components $\zeta_{1}$ and $\zeta_{2}$.

Having analysed the $AdS_{3}$ space we now deal with the BTZ black hole. In global coordinates, it is described by the following metric
\begin{align}
	\dd s^{2} =
	-N^{2} \dd t^{2} 
	+ 
	N^{-2} \dd r^{2}
	+
	r^{2} \left( N^{\phi}\dd t + \dd \phi \right)^{2}
	\ ,
	\label{BTZ1}
\end{align}
where $N$ and $N^{\phi}$ are defined as
\begin{align}
	N = & \sqrt{- M + r^{2} + \frac{J^{2}}{4 r^{2}}}
	\ ,
	& N^{\phi} = & - \frac{J}{ 2 r^{2}} 
	\ ,
	\label{BTZ2}
\end{align}
The parameter $M$ is identified with the mass of the black hole while $J$ represents its angular momentum.
Notice that neither $N$ nor $N^{\phi}$ depends on coordinate $t$ which means that the solution is stationary. 
The non--zero vielbein components  are
\begin{align}
	e^{0} = & N \dd t \ ,
	& 
	e^{1} = & N^{-1} \dd r
	\ , &
	e^{2} = & r N^{\phi} \dd t + r \dd \phi 
	\ , 
	\label{BTZviel}
\end{align}
and the non--zero spin connection components read
\begin{align}
	\omega^{0}{}_{1}= &
	r \dd t - \frac{J}{2 r} \dd \phi
	\ , &
	\omega^{0}{}_{2} = & - \frac{J}{2 r^{2} N} \dd r
	\ , &
	\omega^{1}{}_{2} = & - N \dd \phi
	\ .
	\label{BTZspinC}
\end{align}
The existence of a horizon is constrained by \cite{Banados:1992wn,Banados:1992gq}
\begin{align}
	M&>0 \ , & \left| J\right| & \leq M \ ,
	\label{horiz1}
\end{align}
the case $J=0$, $M=-1$ reduces to empty $AdS_{3}$ (\ref{AdSCglobal1}) while the case $-1 < M < 0$ can be excluded from the physical spectrum since it corresponds to a naked singularity. 
Note also that for $r\rightarrow\infty$ the metric approaches the empty $AdS_{3}$ solution (\ref{AdSCglobal1}). 
It exists also an extremal solution for $M=\left| J \right|$, which preserves two of the four supersymmetries of $AdS_{3}$. In the present case we want to focus on the non--extremal case where all supersymmetries are broken.

It is useful to define the following real bilinears
\begin{align}
	\bB_{0} = &
	- i \zeta^{\dagger}\zeta
	\ ,
	& \bB_{1} = &
	- i \zeta^{\dagger}\sigma_{1}\zeta
	\ ,	
	& \bB_{2} = &
	i\zeta^{\dagger}\sigma_{2}\zeta
	\ ,
	& \bB_{3} =  &
	- i \zeta^{\dagger}\sigma_{3}\zeta
	\ ,
	\label{realBil0}
\end{align}
from which we compute 
\begin{align}
	\bar\epsilon \epsilon 
	& = 
	\bB_{2}
	\ ,\nonumber\\
	\bar\epsilon \Gamma_{0} \epsilon 
	& = 
	i \left[ - f \bB_{0} + r \left( c \bB_{3} -s \bB_{1} \right) \right]
	\ ,\nonumber\\
	\bar\epsilon \Gamma_{1} \epsilon 
	& = 
	- i \left[ c \bB_{1} + s \bB_{3}\right]
	\ ,\nonumber\\	
	\bar\epsilon \Gamma_{2} \epsilon 
	& = 
	i \left[ - r \bB_{0} + f \left( c \bB_{3} -s \bB_{1} \right) \right]
	\ ,
	\label{BTZbilinears1}
\end{align}
where $c=\cos\left( t - \phi \right)$ and $s=\sin\left( t - \phi \right)$.
Due to the anticommutitive nature of $\zeta_{1}$ and $\zeta_{2}$, the following identites hold
\begin{align}
	\bB_{1}^{2}=\bB_{2}^{2}=\bB_{3}^{2}=-\bB_{0}^{2}\,,\quad
	\bB_{i}\bB_{j}=0\,,\, i\neq j
	\,,\quad
	\bB_{0}^{n}=0\,,\, n>2
	\ .
	\label{fierz1}
\end{align}

\section{Wig for General BTZ Black Hole}

The superpartner of a generic field $\Phi$ is constructed by acting with a finite supersymmetry transformation on the original field \cite{Burrington:2004hf}:
\begin{align}
	\boldsymbol{\Phi} = e^{\delta_{\epsilon}} \Phi = \Phi + \delta_{\epsilon} \Phi + \frac{1}{2} \delta_{\epsilon}^{2} \Phi + \dots
	\ .
	\label{defsuperpartner}
\end{align}
In the present case, it is more useful to deal with an expansion in powers of bilinears of $\epsilon$. This is denoted by the superscript $\left[ n \right]$, which counts the number of bilinears. Due to our choice of the background fields, we have
\begin{align}
	B^{\left[ n \right]}&= \frac{1}{2n!}\delta_{\epsilon}^{2 n}B
	\ ,&
	F^{\left[ n \right]}&= \frac{1}{(2n-1)!}\delta_{\epsilon}^{2 n - 1}F
	\ ,&
	n>0 \ ,
	\label{defTransf}
\end{align}
where $B$ and $F$ are respectively bosonic and fermionic fields. Then, for fermionic fields $\left[ n \right]$ counts $n-1$ bilinears plus a spinor $\epsilon$ while for bosonic fields it indicates $n$ bilinears. The $n=0$ case represents the background fields
\begin{align}
	e_{M}^{[0]\,A} & =  \left. e_{M}^{A}\right|_{BTZ} \ ,
	& \psi_{M}^{[0]} = 0 \ ,& 
	& A_{M}^{[0]} = 0 \ .&
	\label{back1}
\end{align}
In this formalism, from susy transformations (\ref{action4}) we derive algorithms to compute iteratively the various fields
\begin{align}
	\psi_{M}^{\left[ n \right]}
	& = \frac{1}{(2n-1)} {\cal D}^{\left[ n \right]}_{M} \epsilon
\ , \nonumber \\
	e_{M}^{\left[ n \right]\,A}
	& = 
	\frac{1}{4 (2n)}  \bar\epsilon \Gamma^{A} \psi_{M}^{\left[ n \right]} + h.c.
\ , \nonumber \\
	A_{M}^{\left[ n \right]}
	& =
	\frac{i}{4(2n)}\bar\epsilon \psi^{\left[ n \right]}_{M} + h.c.
	\ .
	\label{algVielbein}
\end{align}
Then, the wig order by order is written as\footnote{Note that $e_{\left( M \right.}e_{\left. N \right)}=\frac{1}{2}\left( e_{M}e_{N}+e_{N}e_{M} \right)$.}
\begin{align}
	g_{M N}^{\left[ n \right]}
	=  &
	\sum_{p=0}^{n} e^{\left[ p \right]\,A}_{\left( M \right.} \, e^{\left[ n-p \right]\,B}_{\left.  N \right)} \eta_{AB}
	\ .
	\label{algMetric}
\end{align}
At first order in bilinears the gravitino $1$--form reads
\begin{align}
	\psi^{\left[ 1 \right]}
	= &
	\frac{1}{2} \left[ 
	\left( N - f \right) \Gamma_{0}
	- \frac{J}{2 r} \Gamma_{2}
	\right] \epsilon
	\left( \dd t - \dd \phi \right)
	+
	\frac{1}{2} \left( \frac{1}{N} - \frac{1}{f} - \frac{J}{2 r^{2} N} \right)
	\Gamma_{1} \epsilon \, \dd r
	\ .
	\label{grav1}
\end{align}
The first order wig is
\begin{align}
	g^{\left[ 1 \right]} 
	=\  &
	\frac{1}{4}  
	\left[ 
		M - r^{2} + f N
	\right] \bB_{2} \,
	\dd t^{2}
	-
	\frac{1}{8 r^{2} N^{2} f}
	\left[ 
		2 r^{2} \left( N - f \right) + f J
	\right] \bB_{2}	\,
	\dd r^{2} 
	+
	\nonumber\\ &
	-
	\frac{1}{8} 
	\left[ 
	J + 2 M - 2 r^{2} + 2 f N
	\right] \bB_{2}	\,
	\dd t \dd \phi
	+
	\frac{J}{8} \bB_{2} \,
	\dd \phi^2
	\ .
	\label{BTZwig1}
\end{align}
The gauge field is\footnote{$c$ and $s$ are defined as in (\ref{BTZbilinears1})}
\begin{align}
	A^{\left[ 1 \right]}
	=\ &
	\frac{1}{8} \left[ 
	\bB_{0} 
	\left( 
	f \left( N - f \right)
	- \frac{J}{2}
	\right)
	+
	\left( c \bB_{3} -s \bB_{1} \right)
	\left( 
	- r \left( N - f \right)
	+
	\frac{f J}{2 r}
	\right)
	\right]
	\left( \dd t - \dd \phi \right)	
	\nonumber\\
	&
	+
	\frac{1}{8}\left( \frac{1}{N} - \frac{1}{f} - \frac{J}{2 r^{2} N} \right)\left( c \bB_{1} + s \bB_{3}\right)\dd r
	\ .
	\label{BTZgauge1}
\end{align}
The first order of the spin connection is obtained from the vielbein postulate. 
Using standard coordinates transformation, the singularity in $N=0$, being apparent as in the zero order metric (\ref{BTZ1}), can be removed. Notice that all dangerous $1/N$ terms appear along the $\dd r$ component. The pattern repeats itself at the second order.

Iterating the procedure, namely by inserting the first order corrections in (\ref{algVielbein}), we derive the second order results. The gravitino is  
\begin{align}
	\psi^{\left[ 2 \right]}
	& =
	\frac{1}{96 r (r+f)^{3/2}}
	\Big[ 
		\Big(
			(J - 2 r f) \left( \bB_{0} - s \bB_{1} + c \bB_{3} \right)
+  \nonumber \\ & \quad\quad
			-
			\left( 
				r^{2} J + 2 r f (J + r^{2}) + f^{2} (J + 4 r^{2}) + 2 r f^{3} - 2 r N (r + f)^{2}
			\right)
\times	  \nonumber \\ & \quad\quad	\times
			\left( 	 \bB_{0} + s \bB_{1} - c \bB_{3}	\right)
		\Big)
		\big( 
		( \one + \Gamma_{1} ) + ( r + f )( \one - \Gamma_{1} )		
		\big)
+  \nonumber \\ & \quad\quad
		+
		\Big( 
			( - J - 2 r N + 2 r f ) ( \one - \Gamma_{1} )
			+
			( r J - 2 r N (r + f) 
+  \nonumber \\ & \quad\quad			
			+
			f ( J + 2 r^{2} + 2 r f)) ( \one + \Gamma_{1} )
		\Big)
		(r + f) \Gamma_{0} \bB_{2}
	\Big] 
\times \nonumber \\ & \quad \times
	\left( 
		\sin \left[ \frac{t - \phi}{2} \right]  \one - \cos \left[ \frac{t - \phi}{2} \right]  \Gamma_{0} 
	\right)
	\epsilon \, \left( \dd t - \dd \phi \right)
+
\nonumber \\ &
\quad +
	\frac{1}{96 r^{2} N f ( r + f)^{1/2}}
	\left( 2 r^{2} N + f (J - 2 r^{2}) \right)
	\times
\nonumber \\ & \quad\quad
	\times
	\Big( - (\one + \Gamma_{1}) + (r + f) (\one - \Gamma_{1}) \Big) 
	\left( \cos \left[ \frac{t - \phi}{2} \right]  \one - \sin \left[ \frac{t - \phi}{2} \right]  \Gamma_{0} \right) \bB_{2} \epsilon\, \dd r
	\ .
	\label{BTZgrav2}
\end{align}
The second order wig reads
\begin{align}
	g^{\left[ 2 \right]} 
	& =
	\frac{1}{128 r^{2}} 
	\left[ 
		-J^{2} 
		+
		2 r^{2} \left( N - f \right)
		\left( 2 N - f \right)
	\right]
	\bB_{0}^{2}\,\dd t^{2}
	+
	\nonumber\\&
	+
	\frac{1}{256 r^{2}} 
	\left[ 
		 J\left( 3 r^{2} + 2 J \right)
		-
		4 r^{2} 
		\left( N - f \right)
		\left( 3 N - 2 f \right)
	\right]
	\bB_{0}^{2}\,\dd t \dd \phi
	+
	\nonumber\\&
	-
	\frac{1}{256 r^{2} N^{2} f^{2}}
	\left( 
	J f + 2 r^{2}\left( N - f \right)
	\right)
	\left( J f + 2 r^{2}\left( N - 2 f \right) \right)
	\bB_{0}^{2}\,\dd r^{2}
	+
	\nonumber\\&
	-
	\frac{1}{256 r^{2}} 
	\left[ 
		J\left( J + 2 r^{2} \right)
		- 4 r^{2} \left( N - f \right)^{2}
	\right]
	\bB_{0}^{2}\,\dd \phi^{2}
	\ .
	\label{BTZwig2}
\end{align}
The second order gauge field is zero.

We note that first order wig (\ref{BTZwig1}) is proportional to the bilinear $\bB_{2}$ only, while the second order one (\ref{BTZwig2}) depend on $\bB_{0}^{2}=-\bB_{2}^2$. In addiction, for $J\rightarrow0$ and $M\rightarrow-1$ we recover the $AdS_{3}$ solution since in that case the supersymmetry preserves the solution and then there is no wig at all. 
The complete wig does not depend on $t$ and $\phi$, therefore the isometries of the BH are preserved. In the next section we will compute the associated conserved charges, namely the mass and the angular momentum.
The gauge field, which is zero at the bosonic level, is generated at the first order, but it receives no contribution at higher orders.

\section{Conserved Charges}

To investigate the properties of the black hole wig we compute its conserved charges, using holographic technique based on the boundary energy momentum tensor $T_{\mu\nu}$ \cite{Henneaux:1999ib,Gentile:2011jt,Henneaux:1984ei,Brown:1992br,Balasubramanian:1999re}. In the following, Greek indices label boundary directions $t,\phi$. To perform the computation we cast the boundary metric $\gamma_{\mu\nu}$ in ADM--like form
\begin{align}
	\gamma_{\mu\nu} \dd x^{\mu} \dd x^{\nu} 
	& =
	- N_{\Sigma}^{2} \dd t^{2}
	+
	\sigma
	( 	\dd \phi + N^{\phi}_{\Sigma} \dd t )^{2}
	\ ,
	\label{ADM1}
\end{align}
where $\Sigma$ is the $2$--dimensional surface at constant time and the integration is over a circle at spacelike infinity.
The conserved charges associated to the Killing vectors $\xi$ are defined as 
\begin{align}
	Q_{\xi} & =
	\lim_{r\rightarrow\infty}\oint \dd \phi \sqrt{\sigma} u^{\mu} T_{\mu\nu} \xi^{\nu}
	\label{charges1}
\end{align}
where  $u^{\mu}=N_{\Sigma}^{-1}\delta^{\mu t}$ is the timelike unit vector normal to $\Sigma$. 

In the present case, the wig does not depend on $t$ and $\phi$. Thus, the two resulting Killing vectors are 
\begin{align}
	\xi_{1}^{\mu} &= \delta^{\mu t} 
	\ ,
	&\xi_{2}^{\mu} = - \delta^{\mu\phi} 
	\ .&
	\label{charges2}
\end{align}
The associated charges are respectively the mass $M_{tot}$ and the angular momentum $J_{tot}$. After a short computation we find
\begin{align}
	M_{tot} & =
	M 
	+
	\frac{1}{8}\left( 1 + M + J \right)
	\left( <\bB_{2}> - \frac{1}{16} <\bB_{0}^{2}> \right)
	\ , \nonumber \\
	J_{tot} & =
	J 
	+
	\frac{1}{8}\left( 1 + M + J \right)
	\left( <\bB_{2}> - \frac{1}{16} <\bB_{0}^{2}> \right)
	\ .
	\label{MandJ}
\end{align}
The charges ought to be numbers with a given value, then in formula (\ref{MandJ})  the bilinears $\bB_0$ and $\bB_2$  are substituted with their v.e.v. $<\bB^{2}_{0}>$, $<\bB_{2}>$.
In that way the mass and the angular momentum $M_{tot}$ and $J_{tot}$ make sense. A vacuum with non--vanishing v.e.v. for bilinears might explicitly break supersymmetry, leading to a modified mass and angular momentum which depend on them. 

Note that $M_{tot}-J_{tot}=M-J$ and the fermionic corrections do not affect the difference between mass and angular momentum. Thus, if the extremality condition is imposed we expect that it is not lifted \cite{INPREP}.

From action (\ref{action0}) we derive the conserved electric charge $q$
\begin{align}
	q & = 
	\lim_{r\rightarrow\infty}
	\frac{1}{2} \oint \dd \phi \sqrt{\sigma} N_{\sigma}\, \varepsilon^{t M N} i  \bar\psi_{M} \psi_{N}
	\ .
	\label{chargeQ}
\end{align}
Using the equations of motion (\ref{EoMs1}) we can rewrite it in terms of the field strength of gauge field $A$. The computation shows that the leading term of the integral in the large $r$ expansion is $O\left( \frac{1}{r} \right)$, thus in the $r\rightarrow\infty$ limit $q$ vanishes.

The supercharge $\mathcal{Q}$ is connected to the presence of Killing spinors \cite{Belyaev:2007bg,Belyaev:2008ex}. As we have already pointed out, the present work deals with non--extremal BTZ black hole and therefore supersymmetry is totally broken. As a consequence, no Killing spinor exists and thus there is no conserved supercharge.\\

In the present work we construct the complete non linear solution of supergravity $\mathcal{N}=2$, $D=3$ equations of motion starting from the BTZ black hole and generating the fermionic corrections by a finite supersymmetry transformation. 
Due to the simplicity of the framework we are able to give analytic expression for the metric $g_{MN}$ at the highest order in the fermionic expansion.

The aim of this work is to prepare the ground for a complete computation of gravity/fluid correspondence type and derive the non--linear supersymmetric Navier--Stokes equations for the complete solution.
For that we need to replace the fermionic parameters with local function of the boundary of $AdS_{3}$ and insert this variation in the equations of motion. They are satisfied only whether those parameters enjoy certain equations that we want to uncover.
That will be part of a forthcoming publication.

\section*{Acknowledgements}

We thank L. Sommovigo and M. Porrati for useful discussions.

\end{document}